# Genomic tests of variation in inbreeding among individuals and among chromosomes.

Joshua G. Schraiber[1], Stephanie Shih[1], and Montgomery Slatkin[1]

[1] Department of Integrative Biology, University of California, Berkeley, CA 94720-3140, USA

Corresponding author:

Montgomery Slatkin
Department of Integrative Biology
University of California
Berkeley, CA 94720-3140

slatkin@berkeley.edu

FAX: 510-643-6264

**Abstract**

We examine the distribution of heterozygous sites in nine European and nine Yoruban individuals whose genomic sequences were made publicly available by Complete Genomics™ . We show that it is possible to obtain detailed information about inbreeding when a relatively small set of whole-genome sequences is available. Rather than focus on testing for deviations from Hardy-Weinberg genotype frequencies at each site, we analyze the entire distribution of heterozygotes conditioned on the number of copies of the derived (non-chimpanzee) allele. Using Levene's exact test, we reject Hardy-Weinberg in both populations. We generalized Levene's distribution to obtain the exact distribution of the number of heterozygous individuals given that every individual has the same inbreeding coefficient, $F$. We estimated $F$ to be 0.0026 in Europeans and 0.0005 in Yorubans, but we could also reject the hypothesis that $F$ was the same in each individual. We used a composite likelihood method to estimate $F$ in each individual and within each chromosome. Variation in $F$ across chromosomes within individuals was too large to be consistent with sampling effects alone. Furthermore, estimates of $F$ for each chromosome in different populations were not correlated. Our results show how detailed comparisons of population genomic data can be made to theoretical predictions. The application of methods to the Complete Genomics data set shows that the extent of apparent inbreeding varies across chromosomes and across individuals, and estimates of inbreeding coefficients are subject to unexpected levels of variation which might be partly accounted for by selection.

Deviations from Hardy-Weinberg genotypes frequencies (HWF) can have many causes, including population subdivision, natural selection, assortative mating by phenotype, consanguineous matings or the avoidance of consanguineous matings. In general, population subdivision and inbreeding or inbreeding avoidance will affect the whole genome while selection and assortative mating will affect only those loci associated with particular phenotypes. Methods for testing for genome-wide and local deviations from HWF are useful for detecting these processes and for detecting evidence of inbreeding depression resulting from excess homozygosity (CHARLESWORTH and WILLIS 2009). In this paper, we develop and apply methods for using whole-genome sequences to detect small deviations from HWF, both locally and globally, and compare detailed theoretical predictions with publically available sequence data for two human populations. Even small deviations from HWF are detectable at individual loci if sample sizes are large enough. Several methods for testing for deviations from HWF have been developed and tested extensively (reviewed by Weir (1996), ch. 3). The availability of whole-genome sequences from several individuals creates the opportunity to detect systematic deviations from HWF even if sample sizes are small.

In this paper, we show that the distribution of the number of heterozygous individuals, stratified by the number of copies of the derived (i. e. non-chimpanzee) allele, indicates significant deviations from HWF. To illustrate the utility of our approach, we examine two sets of nine genomic sequences made available by Complete Genomics™ (CG) ((http://www.completegenomics.com/public-data/)). These are high-coverage sequences (51X-89X). The error rate for the methodology used is reported to be low (1 miscalled variant per 100 kb) (DRMANAC *et al.* 2010). We compare the distribution of the number of heterozygous sites with the expectations based on the exact distribution derived by Levene (1949) and show we can reject HWF in both data sets. We then test whether we can fit a model in which each individual is inbred to the same extent and reject that model also. Next, we show that there are small but significant differences in the estimated inbreeding coefficient across individuals. Even with that assumption, however, there is more variation in the estimated inbreeding coefficients for different chromosomes than is consistent with sampling effects. We conclude that simple theoretical models do not fit the data closely and that there are additional sources of variation in estimated inbreeding coefficients across chromosomes that are unaccounted for by simple models. We find that in one of the two populations (CEU) there is a slight effect of distance from coding genes, suggesting a subtle role for selection. High-quality genomic sequences provide a detailed picture of deviations from HWF that warrant further exploration.

**Data**

We downloaded the VCF file containing 54 unrelated individuals from Complete Genomics. Although we restricted our later analyses to the eighteen individuals in the CEU and YRI populations (nine in each), we used the larger data set to filter out sites whose genotypes appeared to be the result of alignment errors of duplicated regions. Such regions would be expected to create sites that are apparently heterozygous in every or nearly every individual, thus creating artifactual deviations from HWF. Supplementary Figure 1a shows the distribution of the number of heterozygous individuals in the whole data set and Supplementary Figure 1b focuses on sites with large numbers of heterozygous sites. There is an excess of sites where all 54 individuals are heterozygous.

We chose to restrict to sites at which fewer than 49 of 54 individuals were heterozygous. Using other cutoff values did not significantly change our results. In addition, we filtered out all tri-allelic sites.

**Expectation with no inbreeding**

We assume that a sample of $n$ diploid individuals from a population has been genotyped. Each polymorphic site is characterized by $i$, the number of copies of the derived (i. e. non-chimpanzee) allele, $i=1, \ldots, 2n-1$. Levene (1949) showed that in a randomly mating population, conditional on $i$ and $n$, the probability distribution of $h$, the number of heterozygous individuals, is

$$P(h;i,n) = \Pr(h \mid i,n) = \frac{i!(2n-i)!}{(2n)!} \frac{2^h n!}{((i-h)/2)!h!(n-i-h)/2)!} \qquad (1)$$

$P(h; i, n)$ is unchanged if $i$ is replaced by $2n-i$.

**Single inbreeding coefficient**

We next compare the empirical distribution of $h$ with the distribution expected if every individual has the same positive inbreeding coefficient, $F$. We derive that distribution from Eq. (1) by assuming that, at each polymorphic site, each individual has a probability $F$ of being identical by descent (ibd), and a probability $1-F$ of not being ibd. If an individual is ibd, we assume it is homozygous at that site. That is, we are assuming identity by descent is equivalent to identity in state. The number of individuals that are potentially heterozygous at that site is the difference between $n$ and the number that are ibd. This derivation does not allow for negative values of $F$, which would indicate a significant excess of heterozygotes. Assuming a single positive $F$ for each individual is appropriate when an excess of homozygotes is the result of the mixing of two or more subpopulations (the Wahlund effect).

We assume the distribution of the number of individuals $j$ who are ibd at a site is binomially distributed. If $i$ is even, then that distribution has sample size $n$ and probability $F$. If $i$ is odd, then at least one individual is not ibd because there has to be at least one heterozygous site. In that case, the distribution of $j$ is binomial with sample size $n-1$ and probability $F$. These two cases can be combined by defining $n' = n$ if $i$ is even and $n' = n-1$ if $i$ is odd:

$$\Pr(j \mid n', F) = \binom{n'}{j} F^j (1-F)^{n'-j}. \tag{2}$$

Given $j$, the number $k$ of individuals who are homozygous for the derived allele because they are ibd has a hypergeometric distribution

$$\Pr(k \mid j, i', n') = \frac{\binom{i'/2}{k}\binom{n'-i'/2}{j-k}}{\binom{n'}{j}} \tag{3}$$

where $i' = i$ if $i$ is even and $i' = i-1$ if $i$ is odd. The logic is that there are $i'/2$ pairs of $A$ available to be drawn with replacement from $n'$ pairs.

Given $j$ and $k$, the probability that there are $h$ individuals heterozygous at the site has the distribution given by Eq. (1) with $n=n-j$ and $i=i-2k$. Therefore, summing over $j$ and $k$ yields

$$\Pr(h \mid i, n, F) = \sum_{j=0}^{n'} \sum_{k=0}^{j} P(h, i-2k, n-j) \Pr(j \mid F, n') \Pr(k \mid j, i', n') \tag{4}$$

If $F=0$, Eq. (4) reduces to Eq. (1) because the last two probabilities are 0 except for $j=k=0$.

To use Eq. (4) as a likelihood for each $i$, we replace $n$ by $n_i$, the observed number of sites at which there are $i$ derived alleles, and $h$ by $h_i$, the number of those sites that are heterozygous. It is then straightforward to obtain the maximum likelihood estimate (MLE) of $F$ for that $i$ $\hat{F}_i$. An alternative is to assume the same value of $F$ for all $i$. In that case, the likelihoods from Eq. (4) as a function of $F$ for each $i$ are multiplied to obtain an overall likelihood of $F$, from which the MLE is computed. To test whether a single $F$ is appropriate for all $i$, we used a likelihood ratio test of the hypothesis that all $\hat{F}_i$ are equal. Allowing for $F$ to depend on $i$ lets us test whether lower frequency derived alleles, which will be younger on average, indicate different levels of deviation from HWF, something that would possibly indicate the effects of selection.

We computed the empirical distribution of $h$ for the CG data for both populations, as well as the expectations under the model with no inbreeding, the model with a single $F$ for all individuals, and the model with a different $F$ for each $i$ ($2 \leq i \leq 16$). Supplemental Table 1 shows the results summarized for the nine European (CEU) and nine Yoruban (YRI) individuals. When fitting a single $F$ to all allele frequencies, we find $\hat{F} = .0026$ in CEU and $\hat{F} = .0005$ in YRI. These results suggest there is possibly a small Wahlund effect in CEU data but much less so in the YRI data.

We performed a likelihood ratio test to determine whether assuming a different $F_i$ for each $i$ fit the data better than a single $F$ for all $i$. In both the CEU and YRI populations the tests were highly significant: the likelihood ratios were 566 (p = 0.0) and 188 (p = 0.0), respectively with 14 df.

Once we computed the best fit of $\hat{F}$ under each model, we used a multinomial likelihood ratio test to determine if the expected frequencies fit the data. Briefly, let $\boldsymbol{\pi}_i = (\pi_{i0}, \pi_{i1}, \ldots, \pi_{in})$ be the vector of the expected fraction of sites with $i$ copies of the derived allele that have $h=0, 1, \ldots, n$ heterozygotes under a given model, $\boldsymbol{p}_i = (p_{i0}, p_{i1}, \ldots, p_{in})$ be the observed fraction of sites with $i$ copies of the derived allele that have $0, 1, \ldots, n$ heterozygotes, and $\boldsymbol{h}_i = (h_{i0}, h_{i0}, \ldots, h_{i0})$ be the observed count of sites with $i$ copies of the derived allele that have $0, 1, \ldots, n$ heterozygotes. We computed the test statistic

$$\Lambda = -2 \sum_{i=2}^{2n-2} \sum_{j=0}^{n} h_{ij} \log\left(\pi_{ij} / p_{ij}\right). \tag{5}$$

To compute p-values, we used a parametric bootstrap (KUNSCH 1989). In the CEU population, this results in $\Lambda$ of 2165.685, 1854.912 and 1288.163 for the models that assume $F = 0$, $F_i = F$ for all $i$, and $F_i$ not constrained to be equal respectively. These all correspond to p ≈ 0. In the YRI, the values of $\Lambda$ are 773.375, 757.5121 and 568.4466, respectively, each corresponding to p ≈ 0. Thus, even allowing for a different $F$ for each $i$, the data are not consistent with the assumption that the inbreeding coefficient is the same in each individual.

**Individual-specific inbreeding coefficients**

We next allow each individual to have a different inbreeding coefficient. Let $F_j$ be the inbreeding coefficient for individual $j$ ($j=1,\ldots, n$). It is possible in principle to find the MLE of the set $(F_1, \ldots, F_n)$, given $i$ (the number of copies of the derived allele) and $(h_1, \ldots, h_n)$ where $h_j = (h_{j1}, h_{j2}, \ldots, h_{jL})$ in which $h_{jl} = 1$ if individual $j$ is heterozygous at site $l$ and 0 otherwise. It is computationally difficult, however, to compute the exact likelihood of the *F*'s given the data. Instead we will compute the composite likelihood obtained by assuming that each individual's heterozygosity is determined independently:

$$\Pr(h_1, h_2, ..., h_n | F_1, F_2, ..., F_n, i) = \prod_{j=1}^{n} \prod_{l=1}^{L} \Pr(h_{jl} = 1 | F_j, i_l)^{h_{jl}} \left(1 - \Pr(h_{jl} = 1 | F_j, i_l)\right)^{1-h_{jl}} \quad (6)$$

where *L* is the number of polymorphic sites and

$$\Pr(h_j = 1 | F_j, i) = (1 - F_j) \frac{i(2n - i)}{n(2n - 1)}. \quad (7)$$

In Eq. (7), $F_j$ may be negative. Its lower limit is constrained by *i* and *n*, however, because the right hand side has to be less than 1.

Assuming independence of individuals, we can obtain the MLE $F_j$ for each from Eqs. (6) and (7). We used a block jackknife (KUNSCH 1989) with 5 megabase block sizes to compute standard errors on the estimates of $F_j$. Briefly, we broke the genome up into 5 megabase blocks and computed jackknife replicates by removing each block one at a time and finding the MLE of $F_j$ for the rest of the genome. The results are summarized in Table 1. By the theory of the jackknife, the estimate of $F_j$ divided by its standard error is normally distributed with standard deviation 1 (KUNSCH 1989). We first performed a 2-sided Z-test of the hypothesis that $F_j = 0$ for each *j*. All individuals except for one YRI sample strongly rejected $F_j = 0$. Next, we tested for heterogeneity between individuals in their estimated $F_j$. If two individuals have equal $F_j$, the difference in their estimated $F_j$ should be normally distributed with variance equal to the sum of their squared standard error. We therefore performed a 2-sided test of the null hypothesis that the difference was equal to 0 for every pair of individuals. In the CEU, the p-values ranged from 0 to .38, while in YRI the p-values ranged from 0 to .84.

Our likelihood method for estimating $F$ for each individual differs from the method-of-moments estimator implemented in the widely-used program package PLINK (PURCELL *et al.* 2007). PLINK estimates $F$ by equating the number of homozygous sites in an individual to $nF + (1-F)\left(\sum_{i=1}^{L} 2p_i(1-p_i)\right)$ and solving for $F$, where $L$ is the number of polymorphic sites in each individual and $p_i$ is the allele frequency at the ith site. We implemented this method and found a high correlation between estimated $F$'s for each individual (Spearman's $\rho$=0.96). The MLE estimates did not tend to be larger than the method-of-moments estimates (Wilcoxon test: $p$=0.86 for YRI, $p$=1 for CEU). Keller et al. (2011) found that estimates of individual $F$ based on runs of homozygosity (ROH) were somewhat better than estimates obtained from PLINK for the purpose of detecting inbreeding depression in apparently outbred populations. We did not attempt to implement the Keller et al. ROH method because it was developed for SNP data, not sequence data of the type we analyzed.

**Variation among chromosomes**

For each individual, we estimated $F_j$ separately for each chromosome. Figure 1 shows the results summarized for both populations. We tested whether there is more variation between chromosomes within each individual than can be accounted for by random sampling from that individual's genomic background. For a chromosome with $L$ polymorphic loci, we sampled each site by first sampling its allele frequency from the frequency spectrum and then sampling the status of that site (heterozygous or homozygous) from the individual's background frequency of heterozygous sites for sites with the given allele frequency. We then inferred the $F$ that fit the bootstrapped data best. We performed this test 10000 times for each chromosome in each individual; p-values are shown in Supplementary Table 2. Each population had 198 hypothesis tests, and after a simple Bonferroni correction we found that 141 and 132 tests were significant at the 0.05 level in the CEU and YRI populations respectively As shown, there is substantial variation among the median estimates for each chromosome. Furthermore, median estimates of $F$ per chromosome were not correlated between CEU and YRI (Figure 2), thus ruling out the possibility that there are chromosome-specific factors contributing to the variation among chromosomes we detected.

**Distance from genes**

We tested whether there is an effect on the extent of deviation from HWF on the distance from coding sequences. We stratified sites depending on whether they are more or less than 5 kb from the beginning or end of an annotated coding region and computed the MLE of $F$ for each individual and $i$. We did this by creating an artificial chromosome made by concatenating all regions that are within 5kb from genes (or respectively 5kb away from genes) and computing the likelihood using equation (4) assuming the same $F$ for each $i$. That values shown in Table 2 are the maximum likelihood estimates of $F$. We performed a likelihood ratio test of whether there is a significant effect of stratifying by distance from genes. There is for CEU ($p=3 \times 10^{-6}$) but not for YRI ($p=1$). Thus, there is some evidence that selection might be contributing to variation in the apparent inbreeding coefficients.

**Sequencing error**

Although the error rate in the CG data is low, it is not zero. As long as the probability of sequencing error is independent of whether the error is at a site that is actually heterozygous or homozygous, sequencing error will cause the estimated $F$ to be closer to zero than it actually is. Thus, if the true $F$ is slightly positive, the apparent $F$ will be somewhat smaller. The reason is that the Levene distribution (Eq. 1) is derived under the assumption of equally probable configurations of alleles in $n$ pairs. Randomly relabeling alleles because of sequencing error will not disturb the underlying randomness of the configurations but sites with an actual value of $i$, the number of copies of the derived allele, will appear to have a different $i$. Therefore, if the actual distributions of $h$ for each $i$ fit the Levene distribution, the apparent distribution for each $i$ will also fit. What will change is the frequency spectrum, i. e. the numbers of sites with different $i$.

If $F$ is not zero and the probability of sequencing error does not depend on whether a site is homozygous or heterozygous, then the additional randomness introduced by sequencing error will tend to reduce $F$ slightly because it will tend to restore randomness. Therefore sequencing error of this type will cause estimates of $F$ to be slightly too small, with the difference between the apparent and true values being proportional to the error rate.

If the probability of sequencing error does depend on whether a site is heterozygous or homozygous, then the effect of sequencing error on our estimates of $F$ depends on whether there is a greater tendency to create homozygous or heterozygous sites. We have no information about the error structure of the CG data so we did not attempt to compensate for errors of different kinds.

**Discussion and Conclusion**

We have shown that it is possible to test for deviations from HWF and to quantify those deviations when high-quality genome sequences are available for a relatively small number of individuals. We show that a detailed examination of the numbers of heterozygous sites allows us to test hypotheses that are not testable when only small numbers of polymorphic loci are examined. In particular, we can test whether the apparent inbreeding depends on the frequency of the derived allele or on the distance to coding sequence, and we can test for chromosome-specific or individual-specific differences. Finding dependence on allele frequency or distance to coding sequence indicates that selection might interact with deviations from HWF. Finding differences among individuals provides a way to test for subtle inbreeding depression (KELLER *et al*. 2011).

We used the exact distribution of heterozygous sites conditioned on the number of copies of the derived allele at each polymorphic site, to test for a fit to HWF and to estimate inbreeding coefficients both for the set of individuals considered together and for each individual separately. We applied our method to sequences of nine individuals from each of two populations, Europeans and Yorbans. We found that neither population fits HWF but that the deviation was slightly greater in Europeans than in Yorubans. We also found that inbreeding coefficients estimated for each individual differ significantly. In Europeans, but not in Yorubans, there was a significant effect of distance from coding sequence, suggesting a role for selection.

Although the overall fit of the theory to the data is quite good, there are still unexplained sources of variation. Although variation in the inbreeding coefficient among individuals is to be expected because mating is not perfectly random even in a well mixed population, some of the differences in inbreeding coefficient may reflect past population substructure. Variation in the inbreeding coefficients among chromosomes is more difficult to account for. In humans, linkage disequilibrium decays on a scale of a few hundred kb (REICH *et al*. 2001), making it unlikely that differences among chromosomes reflect accidental correlations in the ancestries of different chromosomes.

Furthermore, it seems unlikely that sequencing error would be important cause of variation among chromosomes. The sequence data we used was based on between 51X-89X coverage which implies a low error rate. Our analysis was based on polymorphic sites that had between 2 and 16 copies of the derived nucleotide in 9 individuals. Sequencing errors that created an apparently single polymorphic nucleotide would have no effect. Sequencing errors at sites that were already polymorphic would have to create an additional copy of a nucleotide that was already present, because we ignored sites at which three or four nucleotides were present. As discussed above, sequencing error that is independent of whether a site is heterozygous or homozygous will tend to reduce $F$ slightly, making our estimates somewhat conservative. Only if sequencing error had a strong tendency to create an excess of homozygous sites would our estimates of $F$ be too large.

We conclude that high quality genome sequences permit detailed examination of the fit to HWF. Although the fit to expectations is good when inbreeding coefficients are different for different individuals, there are still deviations from expectations that are not accounted for.

**Software Resources**

We have posted the script used to carry out the analysis described in this paper at http://cteg.berkeley.edu/software.html.

**Acknowledgements**

This research was supported in part by Training Grant T32-HG00047 and NIH Grant R01-GM40282 to MS.

**Figure captions**

Figure 1. Boxplots of estimates of *F* for each chromosome in both CEU and YRI samples. The top row shows data from the CEU population and the bottom row shows data from the YRI population. The left column shows variation within individuals across chromosomes (individuals 1 through 9 are CEU, 10 through 18 are YRI), while the right column shows variation within chromosome across populations.

Figure 2. Scatter plot of median values of *F* for each chromosome in each population. Numbers in the plot indicate the chromosome (1 through 22). The slope of a best-fit line using Major Axis Regression is not significantly different from 0 ($p = 0.64$).

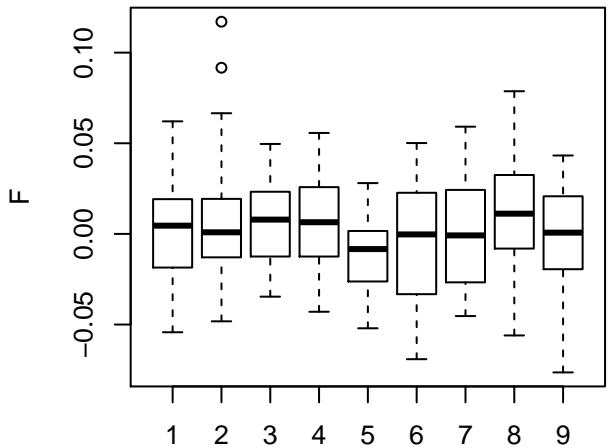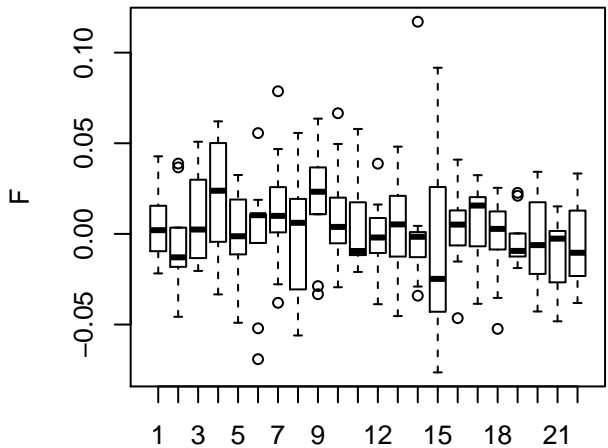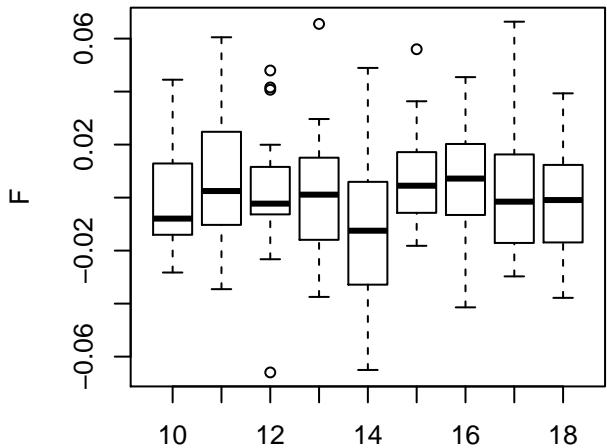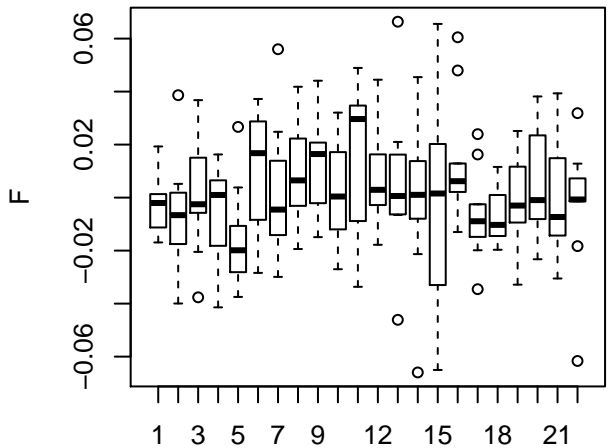

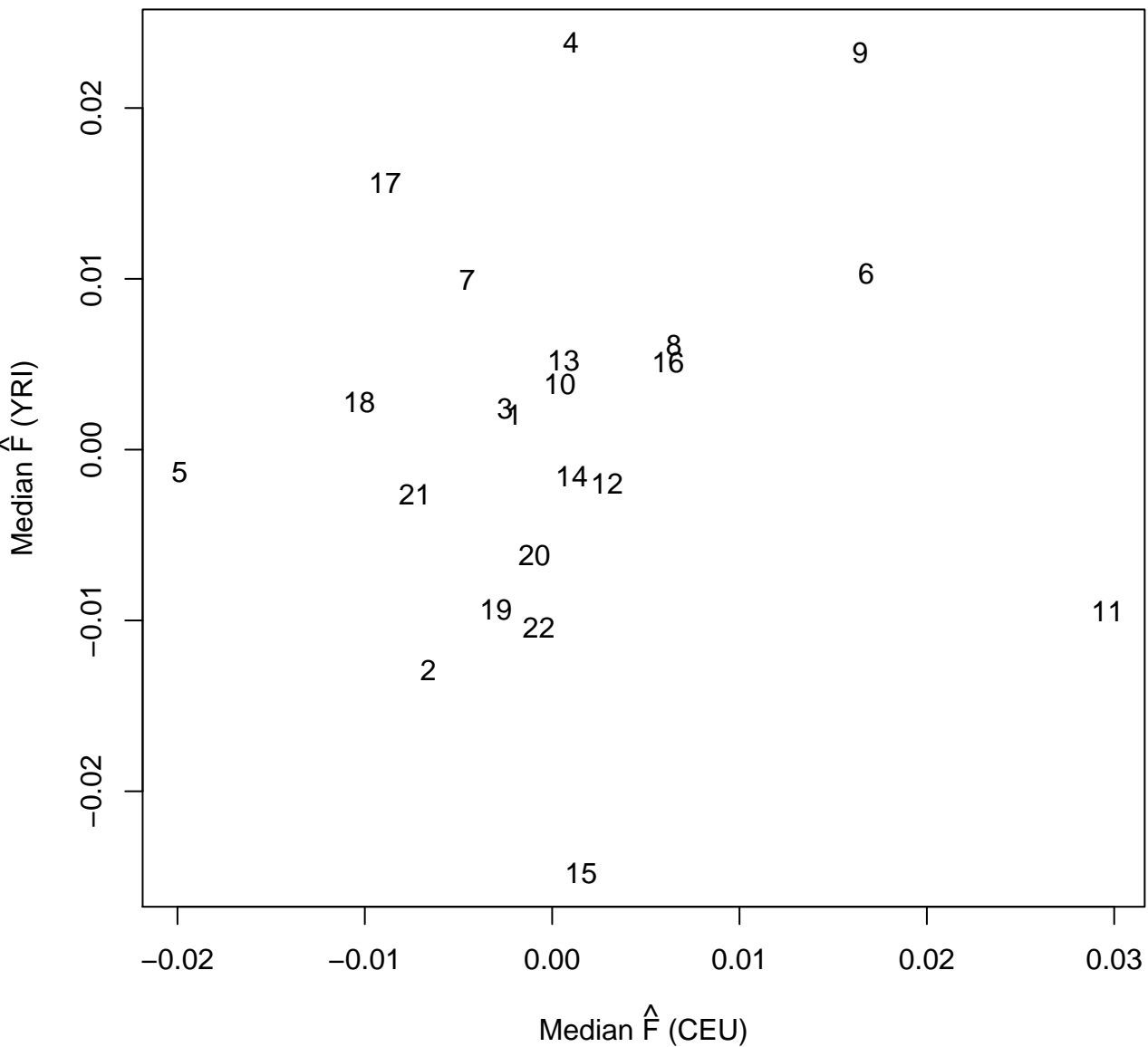